\documentclass[aps,prl,reprint,a4paper,twocolumn,showpacs,floatfix,citeautoscript]{revtex4-1}
\usepackage{graphicx}
\usepackage{amsfonts}
\usepackage{amsmath}
\usepackage{amssymb}
\usepackage{bm}
\usepackage{dcolumn}
\usepackage{hyperref}
\usepackage[usenames]{color}
\hypersetup{pdfborder=0 0 0,colorlinks=true,citecolor=blue,linkcolor=blue}

\begin{document}
\title{Prediction of thickness limits of ideal ultrathin polar films}
\author{Zhicheng Zhong, G. Koster and Paul J. Kelly}
\affiliation{Faculty of Science and Technology and MESA$^+$
Institute for Nanotechnology, University of Twente, P.O. Box 217,
7500 AE Enschede, The Netherlands}
\date{\today}

\begin{abstract}Competition between electronic and atomic reconstruction is a constantly recurring theme in transition metal oxides. We use density functional theory calculations to study this competition for a model system consisting of a thin film of the polar, infinite-layer structure ACuO$_2$ (A = Ca, Sr, Ba) grown on a non-polar, perovskite SrTiO$_3$ substrate. A transition from the bulk planar structure to a chain-type thin film accompanied by substantial changes to the electronic structure is predicted for a SrCuO$_2$ film fewer than five unit cells thick. An analytical model explains why atomic reconstruction becomes more favourable than electronic reconstruction as the film becomes thinner and suggests that similar considerations should be valid for other polar films.
\end{abstract}

%
%
%
%
%
%
%
%
%
%
%
%
%

\maketitle
{\color{red}Introduction:} The lowering of symmetry at surfaces and interfaces frequently results in new or enhanced physical properties. A central goal of transition metal oxide thin film engineering is to exploit this by making interface effects dominate bulk properties in a controlled fashion \cite{Okamoto:nat04,Ohtomo:nat04}. When a polar thin film is grown on a non-polar substrate, achieving this control can be very difficult because an instability arises that can drive electronic and atomic reconstruction. In the case of LaAlO$_3$ (LAO) thin films grown on SrTiO$_3$ (STO) substrates \cite{Ohtomo:nat04,Thiel:sc06,Huijben:am09,Nakagawa:natm06}, the alternate stacking of positively (LaO$^+$) and negatively (AlO$_2^-$) charged layers should result in huge internal electric fields. If nothing else were to happen, the increasing electrostatic energy would give rise to a ``polar electrostatic instability''. The displacement in response to this instability \cite{Nakagawa:natm06,Goniakowski:rpp08} of electrons (charge transfer or leakage) and ions (atomic relaxation and reconstruction) can generate compensating electric fields. Whereas charge transfer and atomic relaxation result in atomically sharp interfaces and ideal thin films with essentially intrinsic properties \cite{Chakhalian:sc07,Junquera:nat03,Zhong:epl08,Popovic:prl08,Pentcheva:prl09,Bristowe:prb09,Chen:prb10}, atomic reconstruction can result in the formation of structures with entirely different intrinsic properties, or in changes to the local stoichiometry due to e.g. oxygen vacancy formation \cite{Herranz:prl07} or ionic intermixing \cite{Kalabukhov:prl09,Chambers:ssr10}. The competition between electronic charge transfer and atomic relaxation and reconstruction in oxide thin films is the source of much puzzling behaviour \cite{Nakagawa:natm06,Herranz:prl07,Kalabukhov:prl09,Chambers:ssr10}.

To unravel the details of this competition is a challenge for experiments because of the limited resolution of interface sensitive measurements and the low visibility of oxygen atoms in most techniques \cite{Nakagawa:natm06}; it is also a challenge for theoretical studies to describe atomic reconstruction that results from subtly different experimental conditions \cite{Chen:prb10,Zhong:prb10} when the role of stoichiometry is unclear. In this paper, we use first-principles total energy calculations to study the growth of thin films of the polar infinite-layer copper oxide ACuO$_2$ (A = Ca, Sr, Ba) on a non-polar perovskite STO substrate and predict a stoichiometric atomic reconstruction as a function of the film thickness. As the parent compound of cuprate high-temperature superconductors (HTS), ACuO$_2$ (ACO) has been intensively studied \cite{Siegrist:nat88,Takano:physc89,Kobayashi:jssc97,Triscone:rpp97}. The recent discovery of very high mobilities in heterostructures containing thin films of SrCuO$_2$ (SCO) \cite{Huijben:arxiv10} and a theoretical proposal for electron-hole liquids \cite{Millis:prb10} has led us to reexamine its structural and electronic properties. Our finding of chain-type formation in thin films as a result of an electrostatic instability and atomic reconstruction provides important insight into the growth of thin films of cuprate HTS. We note that the chain-type formation is different from a GdFeO$_3$-type \cite{Zhong:epl08} or a polar \cite{Pentcheva:prl09} distortion in LAO$|$STO, ferroelectric or antiferrodistortive distortions in PbTiO$_3|$SrTiO$_3$\cite{Bousquet:nat08} and in bulk Ca$_3$Mn$_2$O$_7$\cite{Benedek:prl11}.

{\color{red}Method:} We study ACO thin films in three forms: grown on an STO substrate; in ACO$|$STO multilayers; and freestanding. The thickness of ACO is varied from one to six unit cells while keeping the STO thickness fixed at five unit cells. The in-plane lattice constants of all thin films are fixed at the equilibrium value of the bulk STO substrate calculated to be $a_{\rm STO}=3.87$\AA\ and all atoms are allowed to relax fully. The local density approximation (LDA) calculations were carried out with the projector augmented wave method \cite{Blochl:prb94b} as implemented in the Vienna Ab-initio Simulation Package (VASP) \cite{Kresse:prb99}. Using the generalized gradient approximation (GGA) does not change our main conclusions. To better describe localized Cu $d$ electrons, we use the rotationally invariant LDA+U method \cite{Dudarev:prb98} with $U-J=6.5$~eV \cite{Anisimov:prb91}.

{\color{red}Bulk materials:} Bulk STO is a well-studied band insulator and a popular substrate for growing thin oxide films. It has a typical perovskite structure in which Sr and TiO$_6$ form a CsCl lattice and the TiO$_6$ unit consists of an oxygen octahedron with Ti at its center and the oxygen atoms at the centers of the Sr cube faces. Assigning the formal ionic charges Sr$^{2+}$, Ti$^{4+}$ and O$^{2-}$, it can also be described as an alternate stacking of uncharged SrO$^0$ and TiO$_2^0$ layers \cite{Ohtomo:nat04}.

\begin{figure}[t!]
\includegraphics[width=\columnwidth]{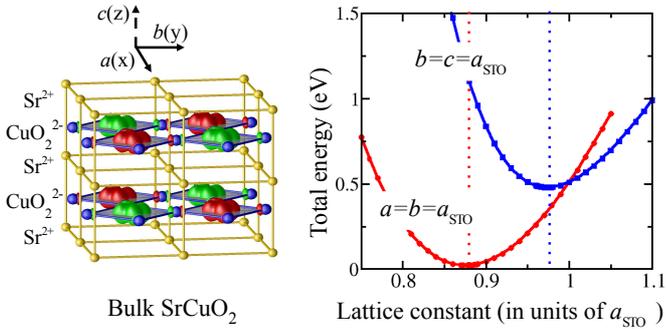}
\caption{Left panel: bulk SrCuO$_2$ with infinite-layer structure. Oxygen atoms are marked by blue spheres. Red and green represent surfaces of constant magnetization density but opposite sign. Right panel: the red curve illustrates the strain energy of bulk SrCuO$_2$ on varying the lattice constant in the $c$ direction keeping the in-plane lattice constants $a$ and $b$ fixed at $a_{\rm STO}$. When the structure is rotated so that $b$ and $c$ are forced to match $a_{\rm STO}$ and $a$ is varied, the resulting strain energy is given by the blue curve.
} \label{bulkstructure}
\end{figure}

The copper oxide ACuO$_2$ with infinite layer structure \cite{Siegrist:nat88} depicted in Fig.~\ref{bulkstructure} can be regarded as a defect perovskite with ordered oxygen vacancies. It consists of positively charged A$^{2+}$ and negatively charged CuO$_2^{2-}$ layers. Because all of the oxygen sites in the AO plane are vacant, there are Cu-O bonds in the $xy$ plane but none in the $z$ direction. The missing atoms and bonds lead to a reduction of the lattice constant in the $z$ direction, reducing the symmetry from cubic in STO to tetragonal in ACO. The lattice constant of ACO increases as the  ion radius increases in the series A = Ca, Sr, Ba. The calculated values given in Table~\ref{lattice} agree well with experiment and SCO is seen to have the smallest lattice mismatch with STO. Though a complex orthorhombic structure for SCO can also be synthesized \cite{Motoyama:prl96}, only the infinite layer structure is taken into consideration in thin film growth experiments \cite{Triscone:rpp97,Koster:jpcm08,Millis:prb10,Huijben:arxiv10}, thus we refer here to the infinite layer structure as the bulk structure and neglect other possibilities in this paper.

\begin{table}[b]
\begin{ruledtabular}
\caption[Tab1]{Lattice constants of perovskite SrTiO$_{3}$ and infinite layer copper oxides ACuO$_{2}$(A=Ca, Sr, Ba).}
\begin{tabular}{l l l l l l l}
& $a^{\rm LDA}$ & $c^{\rm LDA}$  & $a^{\rm GGA}$ &  $c^{\rm GGA}$  &  $a^{\rm exp}$  & $c^{\rm exp}$ \\
\hline
SrTiO$_{3}$ & 3.87 & 3.87 & 3.95 & 3.95 & 3.905                  & 3.905  \\
CaCuO$_{2}$ & 3.77 & 3.08 & 3.87 & 3.20 & 3.853 \footnotemark[1] & 3.177 \footnotemark[1] \\
SrCuO$_{2}$ & 3.84 & 3.38 & 3.95 & 3.47 & 3.926 \footnotemark[2] & 3.432 \footnotemark[2] \\
BaCuO$_{2}$ & 3.92 & 3.68 & 4.03 & 3.84 &  &  \\
\end{tabular}
\label{lattice}
\end{ruledtabular}
\footnotetext[1]{Ref.\onlinecite{Kobayashi:jssc97}}
\footnotetext[2]{Ref.\onlinecite{Takano:physc89}}
\end{table}

Including an on-site Coulomb repulsion term $U$ only slightly affects the calculated lattice constant, and can reproduce the observed antiferromagnetic insulating (AF-I) ground state \cite{Vaknin:prb89} with an energy gap located between filled oxygen $p$ bands and an unfilled minority spin Cu $d_{x^2-y^2}$ band \cite{Anisimov:prb91}. Because its nodes point toward neighbouring oxygen atoms, the $d_{x^2-y^2}$ orbital hybridizes strongly with the oxygen $p$ states and is pushed up in energy. The resulting Cu$^{2+}$ $d^9$ configuration corresponds to all Cu $d$ orbitals being filled except for the minority spin $d_{x^2-y^2}$ orbital. This is apparent in the surfaces of constant magnetization density plotted in Fig.~\ref{bulkstructure} where the spin density on the oxygen atoms also illustrates the AF superexchange coupling between neighbouring Cu ions. The hybridization is mainly confined to the CuO$_2$ plane and does not affect the ionic character of the A$^{2+}$ and CuO$_2^{-2}$ layers so the electrostatic instability of ACuO$_2$ thin films still exists by analogy with LAO thin films.

\begin{figure}[t!]
\includegraphics[width=\columnwidth]{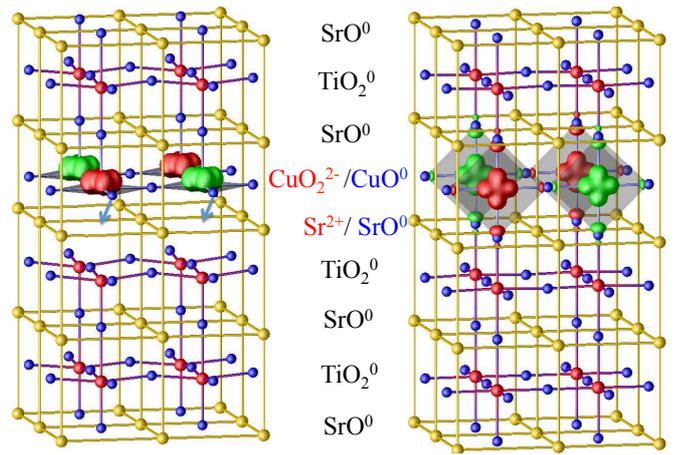}
\caption{Relaxed structures and magnetic isosurfaces of SrCuO$_2|$SrTiO$_3$ multilayers with one unit cell of planar- (left panel) or chain-type (right) SrCuO$_2$ thin film. For simplicity, only three unit cells of SrTiO$_3$ are shown. The arrows in the left panel  schematically indicate the oxygen displacments that transform the planar into a chain-type structure.} 
\label{interfacestructure}
\end{figure}

{\color{red} ACuO$_2$ thin films:} When SCO is grown on an STO substrate, it is assumed that positively charged A$^{2+}$ and negatively charged CuO$_2^{2-}$ layers alternate \cite{Triscone:rpp97}. Because this planar (``CuO$_2$'') structure will lead to an electrostatic instability, we also consider a chain type (``CuO'') structure that is formed by moving one oxygen atom from the CuO$_2^{2-}$ layer to the oxygen vacancy position in the A$^{2+}$ layers as illustrated in Fig.~\ref{interfacestructure} for the case of SCO$|$STO multilayers. This results in a thin film consisting of uncharged SrO$^0$ and chain-type CuO$^0$ layers that does not suffer from an electrostatic instability but at the expense of strain (Fig.~\ref{bulkstructure}). For sufficiently thick films, we expect the planar structure to be most stable. It is not clear {\em a priori} what will occur for thin films of thickness $d$.

\begin{figure}[t!]
\includegraphics[width=\columnwidth]{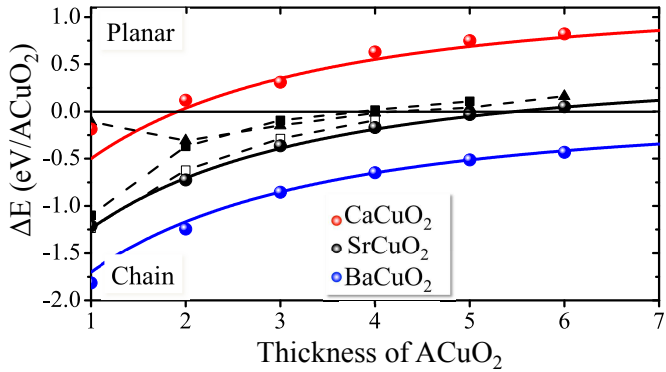}
\caption{Total energy difference between planar and chain-type ACO thin films as a function of the film thickness $d$. Symbols represent LDA results for SrCuO$_2$ thin films in three forms: grown on a SrTiO$_3$ substrate (black circles), as a multilayer with SrTiO$_3$ (squares), and freestanding (triangles); the unfilled squares represent LDA+U results for multilayers. The solid black line for SrCuO$_2$ is based on the analytical model including the electrostatic energy and strain energy discussed in the text while the red and blue solid lines differ only in terms of the calculated strain energies of CaCuO$_2$ and BaCuO$_2$.}
\label{criticalthickness} 
\end{figure}

To answer this question, we calculate the energy difference, $\Delta E=E^{\rm chain}-E^{\rm planar}$ as a function of $d$. As shown in Fig.~\ref{criticalthickness}, $\Delta E$ depends strongly on  both $d$ and on the cation A. For SrCuO$_2$ thin films grown on a lattice matched STO substrate, it increases gradually from -1.2 to +0.05~eV per unit cell, as the thickness increases from one to six unit cells. Since both structures have the same stoichiometry, the sign of $\Delta E$ is a direct measure of their relative stability. Below a critical thickness of about five unit cells, chain type thin films are energetically favourable; above it, planar type films are more stable. In the cation series Ca$\rightarrow$Sr$\rightarrow$Ba, $\Delta E(d)$ decreases but depends on thickness in the same way for all ACuO$_2$ thin films. Similar behavior is found for multilayers and freestanding thin films so surface or interface effects are not essential for a qualitative understanding. We then decompose $\Delta E(d)$ as $ E_C -E_P$, where $E_P$ ($E_p$) denotes the electrostatic energy (density) induced by the instability of planar type thin films and $E_C$ is the chemical bonding energy caused by the oxygen atom moving to form a strained chain-type structure. 

The planar thin films are terminated with positively charged Sr$^{2+}$ and negatively charged CuO$_2^{2-}$ surfaces with nominal areal charge densities $\pm \sigma =\pm e/a^2$ where $a$ is the in-plane lattice constant. They can be modeled as parallel plate capacitors \cite{Nakagawa:natm06,Goniakowski:rpp08,Bristowe:prb09,Chen:prb10,Zhong:prb10,Millis:prb10,Goniakowski:prl07} in which the plates are separated by a thickness $d$ of bulk material with dielectric constant $\varepsilon$.  The electrostatic energy density associated with the electric field $\sigma/\varepsilon\varepsilon_0$ between the plates is $E_p^0 = \sigma^2 / (2\varepsilon\varepsilon_0)$ or approximately $20/\varepsilon$ eV per unit cell, where $\varepsilon_0$ is the dielectric constant of vacuum. This constitutes a strong instability that will drive charge transfer that depends on the thickness and electronic structure of the thin films \cite{Millis:prb10}. If we assume a rigid flat band approximation with constant density of states $\alpha$, the transferred charge is $\sigma / [1+a^2\varepsilon\varepsilon_0/(\alpha d)]$. As a result of the screening by the transferred charge, the residual electrostatic energy density becomes $E_p(d)= E_p^0 / [1+ \alpha d/(a^2\varepsilon\varepsilon_0)]^2$, where the screening factor clearly depends on $\alpha$ and thin film thickness $d$. For a large gap insulator with $\alpha=0$, $E_p$ is a constant \footnote{Note1}. For a good metal with a large $\alpha$, $E_p$ approaches zero. For a small band gap semiconductor with a moderate $\alpha$, $E_p$ vanishes when $d$ is large, consistent with bulk limits; if $d$ is reduced, less charge will be transferred and consequently $E_p$ will be enhanced. When the films are very thin, charge transfer cannot take place to quench the instability.

In that case, $E_P$ can be quenched by displacing oxygen atoms to form a chain-type structure as indicated in Fig.~\ref{interfacestructure}. Because the resulting structure corresponds to a strained ACuO$_2$ thin film grown in a (100) orientation on a (001) STO substrate (rhs of Fig.~\ref{interfacestructure}), rather than with a conventional (001) orientation (lhs of Fig.~\ref{interfacestructure}), doing so will change the bonding energy $E_C$.  The two types of thin film only differ in terms of strain. To estimate the strain energy of planar type thin films, we fix the in-plane lattice constant of bulk ACuO$_2$ at $a_{\rm STO}$ and minimize the total energy with respect to $c$. For chain type thin films, we consider $b$ and $c$ as the in-plane lattice constants, fix them to be equal to $a_{\rm STO}$ and minimize with respect to $a$, as shown in Fig.\ref{bulkstructure}. The energy difference per unit cell between the two minima is $E_C=0.45$~eV for SrCuO$_2$, 1.19~eV for CaCuO$_2$, and -0.01~eV for BaCuO$_2$.

As shown in Fig.~\ref{criticalthickness}, $\Delta E(d)$ can be fit very well using these values for $E_C$ and values of $\alpha=0.1e$/V and $\varepsilon=8$ in the modified parallel plate capacitor model for $E_P$. The very good fit indicates that the competition between the two structures can be represented in terms of electronic and atomic reconstructions and that charge transfer can occur to partly screen the electrostatic instability without altering the atomic structures. However, as the films become thinner, the effect of screening decreases. For fewer than about five unit cells of SCO on STO, the widely assumed planar type thin film cannot be stabilized because of the instability. In spite of a considerable cost in strain energy, the chain type thin film should become energetically favorable because of the lowering of the electrostatic polarization energy $E_P$. 

\begin{figure}[t!]
\includegraphics[width=\columnwidth]{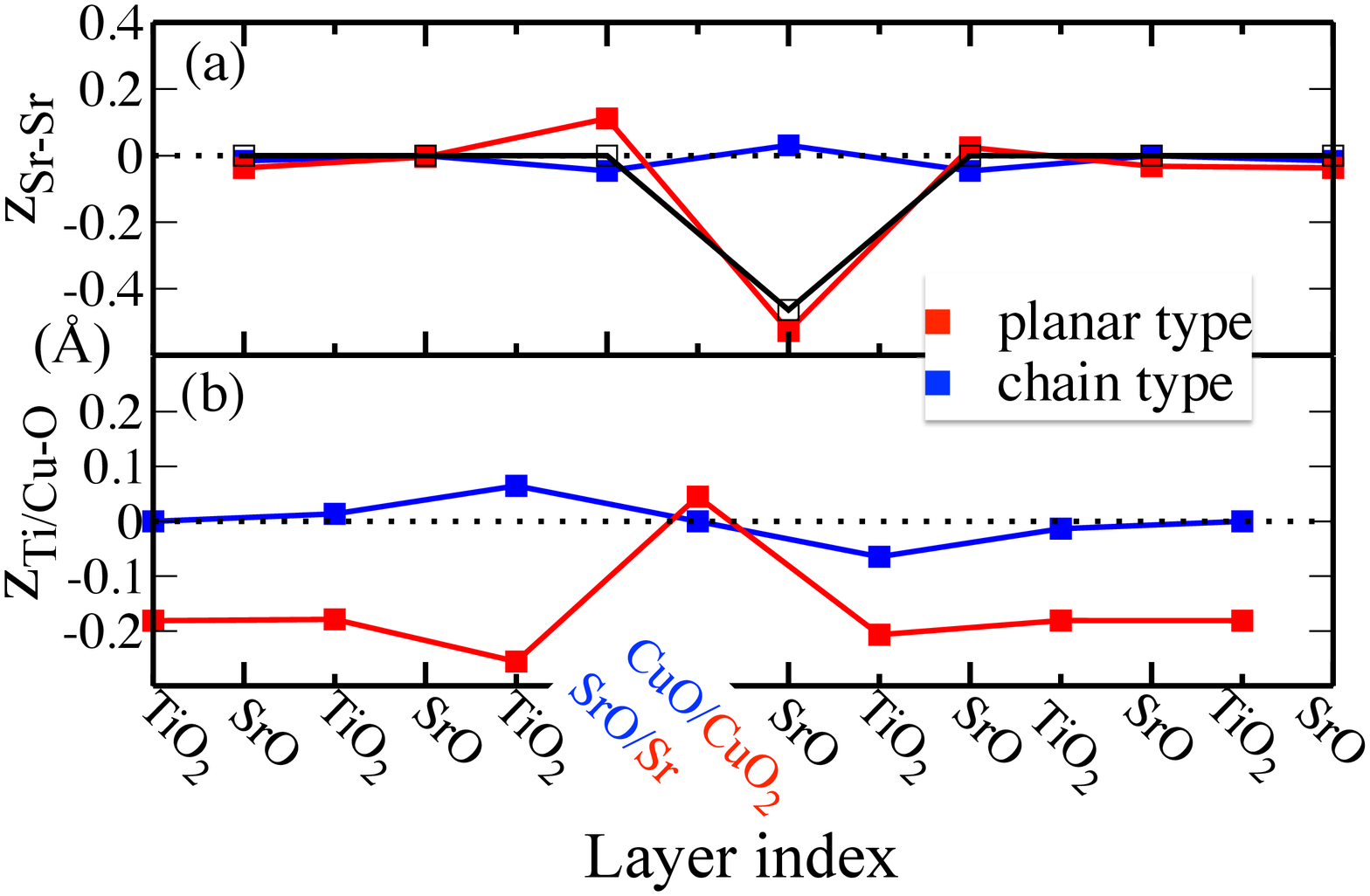}
\caption{(a) Spacings between Sr layers for SCO$|$STO multilayers with planar and chain-type monolayers of SCO relative to $a_{\rm STO}$ for bulk STO. The black squares and line represent ideal stacking without epitaxial strain. (b) The Ti-O or Cu-O corrugation in the $z$-direction for each layer in both types of SCO$|$STO multilayer.} 
\label{interfaceatomic}
\end{figure} 

{\color{red}Atomic and electronic structure:} To discuss the atomic and electronic properties of the novel chain-type thin films, we focus on the multilayers shown in Fig.~\ref{interfacestructure} with a single unit cell of SCO sandwiched between five unit cells of STO. The parallel behaviour of the multilayer and thin film curves (black spheres and squares) in Fig.~\ref{criticalthickness} indicates that the essential physics is the same so we can dispense with discussing the effect of dangling bonds at the free surface. In the chain-type film, there are stiff Cu-O bonds in the growth direction. Because of this, the multilayer containing chain-type SCO is almost 0.5~\AA\ longer in the $z$ direction than that containing planar SCO. This is illustrated in Fig.~\ref{interfaceatomic}(a) where the spacing between layers containing Sr atoms is plotted. Measurement of this spacing should be possible and would provide experimental evidence for chain type thin films. The presence of zigzag Ti-O buckling in Fig.~\ref{interfacestructure} and \ref{interfaceatomic}(b) in the planar-type films is a strong indicator of internal electric fields \cite{Pentcheva:prl09,Bristowe:prb09,Chen:prb10,Zhong:prb10} that should be observable in structural studies. \footnote{note2}

The ground state of the planar-type SCO film is an AF metal. Consistent with a previous LDA+U calculation for YBa$_2$Cu$_3$O$_6$ interfaces with STO \cite{Pavlenko:prb07}, we find that the $d_{x^2-y^2}$ orbital character of the Cu$^{2+}$ $d^9$ minority spin hole is largely unchanged by the polar nature of thin films and the presence of apical oxygen atoms in the STO substrate that leads to Cu-O-Ti type bonding. In contrast, the ground state of the chain-type film is an AF insulator with a localized magnetic moment of 0.6$\mu_B$ on Cu ions and an energy gap of 1.6 eV. The energy gap between filled O $p$ states and an unfilled minority spin Cu $d$ state will decrease on increasing the thickness of the SCO film.The unfilled Cu $d$ orbital character will be $d_{z^2-y^2}$, as shown in Fig.~\ref{interfacestructure}, because the Cu ions are still in a CuO$_4$ square planar configuration surrounded by four oxygen atoms in the $y$-$z$ plane rather than the original, bulk $x$-$y$ plane. Our results suggest an explanation for the $d_{3z^2-r^2}$ orbital observed at the YBa$_2$Cu$_3$O$_7|$LaMnO$_3$ interface~\cite{Chakhalian:sc07}; the electrostatic instability of CuO$_2$ planes \cite{Koster:jpcm08,Millis:prb10} may be driving oxygen atoms (or vacancies) to move and effectively tilt the CuO$_4$ square towards the $z$ direction, favoring orbitals pointing in that direction. The polar nature of CuO$_2$ planes may play a key role in the structure and properties of ultrathin films and interfaces of HTS cuprates.

{\color{red} Conclusion:} Using first-principles calculations, we predict that a polar electrostatic instability will lead to chain type structures being preferred when films of ACuO$_2$ are sufficiently thin and interpret this with a parallel plate capacitor model including charge transfer. As the polar film is made thinner, less charge can be transferred, the residual electrostatic energy increases, and formation of  non-polar chain-type structures becomes an energetically favorable way to resolve the instability without changing the stoichiometry. Similar considerations can be applied to LAO$|$STO. Although pure charge transfer can avoid the ``polar catastrophe''\cite{Nakagawa:natm06,Pentcheva:prl09}, it can not fully quench the electrostatic instability. A large residual electrostatic energy, especially in the low thickness LAO regime, will drive the formation of oxygen vacancies or other defects.

This work was supported by ``NanoNed'', a nanotechnology programme of the Dutch Ministry of Economic Affairs. This work is also part of the research programs of ``Stichting voor
Fundamenteel Onderzoek der Materie'' (FOM) and the use of supercomputer facilities was sponsored by the ``Stichting Nationale Computer Faciliteiten'' (NCF), both financially supported by the ``Nederlandse Organisatie voor Wetenschappelijk Onderzoek'' (NWO).


\begin{thebibliography}{10}%
\makeatletter
\providecommand \@ifxundefined [1]{%
 \ifx #1\undefined \expandafter \@firstoftwo
 \else \expandafter \@secondoftwo
\fi
}%
\providecommand \@ifnum [1]{%
 \ifnum #1\expandafter \@firstoftwo
 \else \expandafter \@secondoftwo
\fi
}%
\providecommand \enquote [1]{``#1''}%
\providecommand \bibnamefont  [1]{#1}%
\providecommand \bibfnamefont [1]{#1}%
\providecommand \citenamefont [1]{#1}%
\providecommand\href[0]{\@sanitize\@href}%
\providecommand\@href[1]{\endgroup\@@startlink{#1}\endgroup\@@href}%
\providecommand\@@href[1]{#1\@@endlink}%
\providecommand \@sanitize [0]{\begingroup\catcode`\&12\catcode`\#12\relax}%
\@ifxundefined \pdfoutput {\@firstoftwo}{%
 \@ifnum{\z@=\pdfoutput}{\@firstoftwo}{\@secondoftwo}%
}{%
 \providecommand\@@startlink[1]{\leavevmode}%
 \providecommand\@@endlink[0]{}%
}{%
 \providecommand\@@startlink[1]{%
  \leavevmode
  \pdfstartlink
   attr{/Border[0 0 1 ]/H/I/C[0 1 1]}%
   user{/Subtype/Link/A<</Type/Action/S/URI/URI(#1)>>}%
  \relax
 }%
 \providecommand\@@endlink[0]{\pdfendlink}%
}%
\providecommand \url  [0]{\begingroup\@sanitize \@url }%
\providecommand \@url [1]{\endgroup\@href {#1}{\urlprefix}}%
\providecommand \urlprefix [0]{URL }%
\providecommand \Eprint[0]{\href }%
\@ifxundefined \urlstyle {%
  \providecommand \doi [1]{doi:\discretionary{}{}{}#1}%
}{%
  \providecommand \doi [0]{doi:\discretionary{}{}{}\begingroup
  \urlstyle{rm}\Url }%
}%
\providecommand \doibase [0]{http://dx.doi.org/}%
\providecommand \Doi[1]{\href{\doibase#1}}%
\providecommand \bibAnnote [3]{%
  \BibitemShut{#1}%
  \begin{quotation}\noindent
    \textsc{Key:}\ #2\\\textsc{Annotation:}\ #3%
  \end{quotation}%
}%
\providecommand \bibAnnoteFile [2]{%
  \IfFileExists{#2}{\bibAnnote {#1} {#2} {\input{#2}}}{}%
}%
\providecommand \typeout [0]{\immediate \write \m@ne }%
\providecommand \selectlanguage [0]{\@gobble}%
\providecommand \bibinfo [0]{\@secondoftwo}%
\providecommand \bibfield [0]{\@secondoftwo}%
\providecommand \translation [1]{[#1]}%
\providecommand \BibitemOpen[0]{}%
\providecommand \bibitemStop [0]{}%
\providecommand \bibitemNoStop [0]{.\EOS\space}%
\providecommand \EOS [0]{\spacefactor3000\relax}%
\providecommand \BibitemShut [1]{\csname bibitem#1\endcsname}%
\bibitem{Okamoto:nat04}%
  \BibitemOpen
  \bibfield{author}{%
  \bibinfo {author} {\bibfnamefont{S.}~\bibnamefont{Okamoto}}\ and\ \bibinfo
  {author} {\bibfnamefont{A.~J.}\ \bibnamefont{Millis}},\ }%
  \bibfield{journal}{%
  \Doi{10.1038/nature02450}{\bibinfo {journal} {Nature}}\ }%
  \textbf{\bibinfo {volume} {428}},\ \bibinfo {pages} {630} (\bibinfo {year}
  {2004})%
  \bibAnnoteFile{NoStop}{Okamoto:nat04}%
\bibitem{Ohtomo:nat04}%
  \BibitemOpen
  \bibfield{author}{%
  \bibinfo {author} {\bibfnamefont{A.}~\bibnamefont{Ohtomo}}\ and\ \bibinfo
  {author} {\bibfnamefont{H.~Y.}\ \bibnamefont{Hwang}},\ }%
  \bibfield{journal}{%
  \Doi{10.1038/nature02308}{\bibinfo {journal} {Nature}}\ }%
  \textbf{\bibinfo {volume} {427}},\ \bibinfo {pages} {423} (\bibinfo {year}
  {2004})%
  \bibAnnoteFile{NoStop}{Ohtomo:nat04}%
\bibitem{Thiel:sc06}%
  \BibitemOpen
  \bibfield{author}{%
  \bibinfo {author} {\bibfnamefont{S.}~\bibnamefont{Thiel}}, \bibinfo {author}
  {\bibfnamefont{G.}~\bibnamefont{Hammerl}}, \bibinfo {author}
  {\bibfnamefont{A.}~\bibnamefont{Schmehl}}, \bibinfo {author}
  {\bibfnamefont{C.~W.}\ \bibnamefont{Schneider}},\ and\ \bibinfo {author}
  {\bibfnamefont{J.}~\bibnamefont{Mannhart}},\ }%
  \bibfield{journal}{%
  \Doi{10.1126/science.1131091}{\bibinfo {journal} {Science}}\ }%
  \textbf{\bibinfo {volume} {313}},\ \bibinfo {pages} {1942} (\bibinfo {year}
  {2006})%
  \bibAnnoteFile{NoStop}{Thiel:sc06}%
\bibitem{Huijben:am09}%
  \BibitemOpen
  \bibfield{author}{%
  \bibinfo {author} {\bibfnamefont{M.}~\bibnamefont{Huijben}}, \bibinfo
  {author} {\bibfnamefont{A.}~\bibnamefont{Brinkman}}, \bibinfo {author}
  {\bibfnamefont{G.}~\bibnamefont{Koster}} {\it et al.}, \ }%
  \bibfield{journal}{%
  \Doi{10.1002/adma.200801448}{\bibinfo {journal} {Adv. Mat.}}\ }%
  \textbf{\bibinfo {volume} {21}},\ \bibinfo {pages} {1665} (\bibinfo {year}
  {2009})%
  \bibAnnoteFile{NoStop}{Huijben:am09}%
\bibitem{Nakagawa:natm06}%
  \BibitemOpen
  \bibfield{author}{%
  \bibinfo {author} {\bibfnamefont{N.}~\bibnamefont{Nakagawa}}, \bibinfo
  {author} {\bibfnamefont{H.~Y.}\ \bibnamefont{Hwang}},\ and\ \bibinfo {author}
  {\bibfnamefont{D.~A.}\ \bibnamefont{Muller}},\ }%
  \bibfield{journal}{%
  \Doi{10.1038/nmat1569}{\bibinfo {journal} {Nature Materials}}\ }%
  \textbf{\bibinfo {volume} {5}},\ \bibinfo {pages} {204} (\bibinfo {year}
  {2006})%
  \bibAnnoteFile{NoStop}{Nakagawa:natm06}%
\bibitem{Goniakowski:rpp08}%
  \BibitemOpen
  \bibfield{author}{%
  \bibinfo {author} {\bibfnamefont{J.}~\bibnamefont{Goniakowski}}, \bibinfo
  {author} {\bibfnamefont{F.}~\bibnamefont{Finocchi}},\ and\ \bibinfo {author}
  {\bibfnamefont{C.}~\bibnamefont{Noguera}},\ }%
  \bibfield{journal}{%
  \Doi{10.1088/0034-4885/71/1/016501}{\bibinfo {journal} {Rep. Prog. Phys.}}\
  }%
  \textbf{\bibinfo {volume} {71}},\ \bibinfo {pages} {016501} (\bibinfo {year}
  {2008})%
  \bibAnnoteFile{NoStop}{Goniakowski:rpp08}%
\bibitem{Chakhalian:sc07}%
  \BibitemOpen
  \bibfield{author}{%
  \bibinfo {author} {\bibfnamefont{J.}~\bibnamefont{Chakhalian}}, \bibinfo
  {author} {\bibfnamefont{J.~W.}\ \bibnamefont{Freeland}}, \bibinfo {author}
  {\bibfnamefont{H.-U.}\ \bibnamefont{Habermeier}} {\it et al.},\ }%
  \bibfield{journal}{%
  \Doi{10.1126/science.1149338}{\bibinfo {journal} {Science}}\ }%
  \textbf{\bibinfo {volume} {318}},\ \bibinfo {pages} {1115} (\bibinfo {year}
  {2007})%
  \bibAnnoteFile{NoStop}{Chakhalian:sc07}%
\bibitem{Junquera:nat03}%
  \BibitemOpen
  \bibfield{author}{%
  \bibinfo {author} {\bibfnamefont{J.}~\bibnamefont{Junquera}}\ and\ \bibinfo
  {author} {\bibfnamefont{P.}~\bibnamefont{Ghosez}},\ }%
  \bibfield{journal}{%
  \Doi{10.1038/nature01501}{\bibinfo {journal} {Nature}}\ }%
  \textbf{\bibinfo {volume} {422}},\ \bibinfo {pages} {506} (\bibinfo {year}
  {2003})%
  \bibAnnoteFile{NoStop}{Junquera:nat03}%
\bibitem{Zhong:epl08}%
  \BibitemOpen
  \bibfield{author}{%
  \bibinfo {author} {\bibfnamefont{Z.}~\bibnamefont{Zhong}}\ and\ \bibinfo
  {author} {\bibfnamefont{P.~J.}\ \bibnamefont{Kelly}},\ }%
  \bibfield{journal}{%
  \Doi{10.1209/0295-5075/84/27001}{\bibinfo {journal} {Europhys. Lett.}}\ }%
  \textbf{\bibinfo {volume} {84}},\ \bibinfo {pages} {27001} (\bibinfo {year}
  {2008})%
  \bibAnnoteFile{NoStop}{Zhong:epl08}%
  \bibitem{Popovic:prl08}%
  \BibitemOpen
  \bibfield{author}{%
  \bibinfo {author} {\bibfnamefont{Z.~S.}\ \bibnamefont{Popovi{\'{c}}}},
  \bibinfo {author} {\bibfnamefont{S.}~\bibnamefont{Satpathy}},\ and\ \bibinfo
  {author} {\bibfnamefont{R.~M.}\ \bibnamefont{Martin}},\ }%
  \bibfield{journal}{%
  \Doi{10.1103/PhysRevLett.101.256801}{\bibinfo {journal} {Phys. Rev. Lett.}}\
  }%
  \textbf{\bibinfo {volume} {101}},\ \bibinfo {pages} {256801} (\bibinfo {year}
  {2008})%
  \bibAnnoteFile{NoStop}{Popovic:prl08}%
\bibitem{Pentcheva:prl09}%
  \BibitemOpen
  \bibfield{author}{%
  \bibinfo {author} {\bibfnamefont{R.}~\bibnamefont{Pentcheva}}\ and\ \bibinfo
  {author} {\bibfnamefont{W.~E.}\ \bibnamefont{Pickett}},\ }%
  \bibfield{journal}{%
  \Doi{10.1103/PhysRevLett.102.107602}{\bibinfo {journal} {Phys. Rev. Lett.}}\
  }%
  \textbf{\bibinfo {volume} {102}},\ \bibinfo {pages} {107602} (\bibinfo
  {month} {Mar}\ \bibinfo {year} {2009})%
  \bibAnnoteFile{NoStop}{Pentcheva:prl09}%
  \bibitem{Bristowe:prb09}%
  \BibitemOpen
  \bibfield{author}{%
  \bibinfo {author} {\bibfnamefont{N.~C.}\ \bibnamefont{Bristowe}}, \bibinfo
  {author} {\bibfnamefont{E.}~\bibnamefont{Artacho}},\ and\ \bibinfo {author}
  {\bibfnamefont{P.~B.}\ \bibnamefont{Littlewood}},\ }%
  \bibfield{journal}{%
  \Doi{10.1103/PhysRevB.80.045425}{\bibinfo {journal} {Phys. Rev. B}}\ }%
  \textbf{\bibinfo {volume} {80}},\ \bibinfo {pages} {045425} (\bibinfo {year}
  {2009})%
  \bibAnnoteFile{NoStop}{Bristowe:prb09}%
\bibitem{Chen:prb10}%
  \BibitemOpen
  \bibfield{author}{%
  \bibinfo {author} {\bibfnamefont{H.}~\bibnamefont{Chen}}, \bibinfo {author}
  {\bibfnamefont{A.}~\bibnamefont{Kolpak}},\ and\ \bibinfo {author}
  {\bibfnamefont{S.}~\bibnamefont{{Ismail-Beigi}}},\ }%
  \bibfield{journal}{%
  \Doi{10.1103/PhysRevB.82.085430}{\bibinfo {journal} {Phys. Rev. B}}\ }%
  \textbf{\bibinfo {volume} {82}},\ \bibinfo {pages} {085430} (\bibinfo {year}
  {2010})%
  \bibAnnoteFile{NoStop}{Chen:prb10}%
\bibitem{Herranz:prl07}%
  \BibitemOpen
  \bibfield{author}{%
  \bibinfo {author} {\bibfnamefont{G.}~\bibnamefont{Herranz}}, \bibinfo
  {author} {\bibfnamefont{M.}~\bibnamefont{Basleti\'{c}}}, \bibinfo {author}
  {\bibfnamefont{M.}~\bibnamefont{Bibes}} {\it et al.}, \ }%
  \bibfield{journal}{%
  \Doi{10.1103/PhysRevLett.98.216803}{\bibinfo {journal} {Phys. Rev. Lett.}}\
  }%
  \textbf{\bibinfo {volume} {98}},\ \bibinfo {pages} {216803} (\bibinfo {year}
  {2007})%
  \bibAnnoteFile{NoStop}{Herranz:prl07}%
\bibitem{Kalabukhov:prl09}%
  \BibitemOpen
  \bibfield{author}{%
  \bibinfo {author} {\bibfnamefont{A.~S.}\ \bibnamefont{Kalabukhov}}, \bibinfo
  {author} {\bibfnamefont{Y.~A.}\ \bibnamefont{Boikov}}, \bibinfo {author}
  {\bibfnamefont{I.~T.}\ \bibnamefont{Serenkov}} {\it et al.}, \ }%
  \bibfield{journal}{%
  \Doi{10.1103/PhysRevLett.103.146101}{\bibinfo {journal} {Phys. Rev. Lett.}}\
  }%
  \textbf{\bibinfo {volume} {103}},\ \bibinfo {pages} {146101} (\bibinfo {year}
  {2009})%
  \bibAnnoteFile{NoStop}{Kalabukhov:prl09}%
\bibitem{Chambers:ssr10}%
  \BibitemOpen
  \bibfield{author}{%
  \bibinfo {author} {\bibfnamefont{S.~A.}\ \bibnamefont{Chambers}}, \bibinfo
  {author} {\bibfnamefont{M.~H.}\ \bibnamefont{Engelhard}}, \bibinfo {author}
  {\bibfnamefont{V.}\ \bibnamefont{Shutthanandan}} {\it et al.}, \ }%
  \bibfield{journal}{%
  \Doi{10.1016/j.surfrep.2010.09.001}{\bibinfo {journal} {Surface Science Reports}}\
  }%
  \textbf{\bibinfo {volume} {65}},\ \bibinfo {pages} {317} (\bibinfo {year}
  {2010})%
  \bibAnnoteFile{NoStop}{Chambers:ssr10}%
\bibitem{Zhong:prb10}%
  \BibitemOpen
  \bibfield{author}{%
  \bibinfo {author} {\bibfnamefont{Z.}~\bibnamefont{Zhong}}, \bibinfo {author} {\bibfnamefont{P.~X.}~\bibnamefont{Xu}}\ and\ \bibinfo
  {author} {\bibfnamefont{P.~J.}\ \bibnamefont{Kelly}},\ }%
  \bibfield{journal}{%
  \Doi{10.1103/PhysRevB.82.165127}{\bibinfo {journal} {Phys. Rev. B}}\ }%
  \textbf{\bibinfo {volume} {82}},\ \bibinfo {pages} {165127} (\bibinfo {year}
  {2010})%
  \bibAnnoteFile{NoStop}{Zhong:prb10}%
\bibitem{Siegrist:nat88}%
  \BibitemOpen
  \bibfield{author}{%
  \bibinfo {author} {\bibfnamefont{T.}~\bibnamefont{Siegrist}}, \bibinfo
  {author} {\bibfnamefont{S.~M.}\ \bibnamefont{Zahurak}}, \bibinfo {author}
  {\bibfnamefont{D.~W.}\ \bibnamefont{Murphy}},\ and\ \bibinfo {author}
  {\bibfnamefont{R.~S.}\ \bibnamefont{Roth}},\ }%
  \bibfield{journal}{%
  \Doi{10.1038/334231a0}{\bibinfo {journal} {Nature}}\ }%
  \textbf{\bibinfo {volume} {334}},\ \bibinfo {pages} {231} (\bibinfo {year}
  {1988})%
  \bibAnnoteFile{NoStop}{Siegrist:nat88}%
\bibitem{Takano:physc89}%
  \BibitemOpen
  \bibfield{author}{%
  \bibinfo {author} {\bibfnamefont{M.}~\bibnamefont{Takano}}, \bibinfo {author}
  {\bibfnamefont{Y.}~\bibnamefont{Takeda}}, \bibinfo {author}
  {\bibfnamefont{H.}~\bibnamefont{Okada}}, \bibinfo {author}
  {\bibfnamefont{M.}~\bibnamefont{Miyamoto}},\ and\ \bibinfo {author}
  {\bibfnamefont{T.}~\bibnamefont{Kusaka}},\ }%
  \bibfield{journal}{%
  \Doi{10.1016/S0921-4534(89)80007-3}{\bibinfo {journal} {Physica C}}\ }%
  \textbf{\bibinfo {volume} {159}},\ \bibinfo {pages} {375} (\bibinfo {year}
  {1989})%
  \bibAnnoteFile{NoStop}{Takano:physc89}%
\bibitem{Kobayashi:jssc97}%
  \BibitemOpen
  \bibfield{author}{%
  \bibinfo {author} {\bibfnamefont{N.}~\bibnamefont{Kobayashi}}, \bibinfo
  {author} {\bibfnamefont{Z.}~\bibnamefont{Hiroi}},\ and\ \bibinfo {author}
  {\bibfnamefont{M.}~\bibnamefont{Takano}},\ }%
  \bibfield{journal}{%
  \Doi{10.1006/jssc.1997.7442}{\bibinfo {journal} {J. Solid State Chem.}}\ }%
  \textbf{\bibinfo {volume} {132}},\ \bibinfo {pages} {274} (\bibinfo {year}
  {1997})%
  \bibAnnoteFile{NoStop}{Kobayashi:jssc97}%
\bibitem{Triscone:rpp97}%
  \BibitemOpen
  \bibfield{author}{%
  \bibinfo {author} {\bibfnamefont{J.-M.}\ \bibnamefont{Triscone}}\ and\
  \bibinfo {author} {\bibfnamefont{O.}~\bibnamefont{Fischer}},\ }%
  \bibfield{journal}{%
  \Doi{10.1088/0034-4885/60/12/004}{\bibinfo {journal} {Rep. Prog. Phys.}}\ }%
  \textbf{\bibinfo {volume} {60}},\ \bibinfo {pages} {1673} (\bibinfo {year}
  {1997})%
  \bibAnnoteFile{NoStop}{Triscone:rpp97}%
\bibitem{Huijben:arxiv10}%
  \BibitemOpen
  \bibfield{author}{%
  \bibinfo {author} {\bibfnamefont{M.}~\bibnamefont{Huijben}} {\it et al.}, \ }%
  \bibinfo {journal} {e-print arXiv:cond-mat/1008.1896v1}%
  \bibAnnoteFile{NoStop}{Huijben:arxiv10}%
\bibitem{Millis:prb10}%
  \BibitemOpen
\bibfield{journal}{%
    }%
  \bibfield{author}{%
  \bibinfo {author} {\bibfnamefont{A.~J.}\ \bibnamefont{Millis}}\ and\ \bibinfo
  {author} {\bibfnamefont{D.~G.}\ \bibnamefont{Schlom}},\ }%
  \bibfield{journal}{%
  \Doi{10.1103/PhysRevB.82.073101}{\bibinfo {journal} {Phys. Rev. B}}\ }%
  \textbf{\bibinfo {volume} {82}},\ \bibinfo {pages} {073101} (\bibinfo {year}
  {2010})%
  \bibAnnoteFile{NoStop}{Millis:prb10}%
  \bibitem{Bousquet:nat08}%
  \BibitemOpen
  \bibfield{author}{%
  \bibinfo {author} {\bibfnamefont{E.}~\bibnamefont{Bousquet}}, \bibinfo
  {author} {\bibfnamefont{M.}~\bibnamefont{Dawber}}, \bibinfo {author}
  {\bibfnamefont{N.}~\bibnamefont{Stucki}} {\it et al.},\ }%
  \bibfield{journal}{%
  \bibinfo {journal} {Nature}\ }%
  \textbf{\bibinfo {volume} {452}},\ \bibinfo {pages} {732} (\bibinfo {year}
  {2008})%
  \bibAnnoteFile{NoStop}{Bousquet:nat08}%
\bibitem{Benedek:prl11}%
  \BibitemOpen
  \bibfield{author}{%
  \bibinfo {author} {\bibfnamefont{N.~A.}\ \bibnamefont{Benedek}}\ and\
  \bibinfo {author} {\bibfnamefont{C.~J.}\ \bibnamefont{Fennie}},\ }%
  \bibfield{journal}{%
  \Doi{10.1103/PhysRevLett.106.107204}{\bibinfo {journal} {Phys. Rev. Lett.}}\
  }%
  \textbf{\bibinfo {volume} {106}},\ \bibinfo {pages} {107204} (\bibinfo {year} {2011})%
  \bibAnnoteFile{NoStop}{Benedek:prl11}%
\bibitem{Blochl:prb94b}%
  \BibitemOpen
  \bibfield{author}{%
  \bibinfo {author} {\bibfnamefont{P.~E.}\ \bibnamefont{Bl{\"{o}}chl}},\ }%
  \bibfield{journal}{%
  \Doi{10.1103/PhysRevB.49.16223}{\bibinfo {journal} {Phys. Rev. B}}\ }%
  \textbf{\bibinfo {volume} {50}},\ \bibinfo {pages} {17953} (\bibinfo {year}
  {1994})%
  \bibAnnoteFile{NoStop}{Blochl:prb94b}%
\bibitem{Kresse:prb99}%
  \BibitemOpen
  \bibfield{author}{%
  \bibinfo {author} {\bibfnamefont{G.}~\bibnamefont{Kresse}}\ and\ \bibinfo
  {author} {\bibfnamefont{D.}~\bibnamefont{Joubert}},\ }%
  \bibfield{journal}{%
  \Doi{10.1103/PhysRevB.59.1758}{\bibinfo {journal} {Phys. Rev. B}}\ }%
  \textbf{\bibinfo {volume} {59}},\ \bibinfo {pages} {1758} (\bibinfo {year}
  {1999})%
  \bibAnnoteFile{NoStop}{Kresse:prb99}%
\bibitem{Dudarev:prb98}%
  \BibitemOpen
  \bibfield{author}{%
  \bibinfo {author} {\bibfnamefont{S.~L.}\ \bibnamefont{Dudarev}}, \bibinfo
  {author} {\bibfnamefont{G.~A.}\ \bibnamefont{Botton}}, \bibinfo {author}
  {\bibfnamefont{S.~Y.}\ \bibnamefont{Savrasov}} {\it et al.}, \ }%
  \bibfield{journal}{%
  \Doi{10.1103/PhysRevB.57.150}{\bibinfo {journal} {Phys. Rev. B}}\ }%
  \textbf{\bibinfo {volume} {57}},\ \bibinfo {pages} {1505} (\bibinfo {year}
  {1998})%
  \bibAnnoteFile{NoStop}{Dudarev:prb98}%
\bibitem{Anisimov:prb91}%
  \BibitemOpen
  \bibfield{author}{%
  \bibinfo {author} {\bibfnamefont{V.~I.}\ \bibnamefont{Anisimov}}, \bibinfo
  {author} {\bibfnamefont{J.}~\bibnamefont{Zaanen}},\ and\ \bibinfo {author}
  {\bibfnamefont{O.~K.}\ \bibnamefont{Andersen}},\ }%
  \bibfield{journal}{%
  \Doi{10.1103/PhysRevB.44.943}{\bibinfo {journal} {Phys. Rev. B}}\ }%
  \textbf{\bibinfo {volume} {44}},\ \bibinfo {pages} {943} (\bibinfo {year}
  {1991})%
  \bibAnnoteFile{NoStop}{Anisimov:prb91}%
\bibitem{Motoyama:prl96}%
  \BibitemOpen
  \bibfield{author}{%
  \bibinfo {author} {\bibfnamefont{N.}~\bibnamefont{Motoyama}}, \bibinfo
  {author} {\bibfnamefont{H.}~\bibnamefont{Eisaki}},\ and\ \bibinfo {author}
  {\bibfnamefont{S.}~\bibnamefont{Uchida}},\ }%
  \bibfield{journal}{%
  \Doi{10.1103/PhysRevLett.76.3212}{\bibinfo {journal} {Phys. Rev. Lett.}}\ }%
  \textbf{\bibinfo {volume} {76}},\ \bibinfo {pages} {3212} (\bibinfo {year}
  {1996})%
  \bibAnnoteFile{NoStop}{Motoyama:prl96}%
\bibitem{Koster:jpcm08}%
  \BibitemOpen
  \bibfield{author}{%
  \bibinfo {author} {\bibfnamefont{G.}~\bibnamefont{Koster}}, \bibinfo {author}
  {\bibfnamefont{A.}~\bibnamefont{Brinkman}}, \bibinfo {author}
  {\bibfnamefont{H.}~\bibnamefont{Hilgenkamp}} {\it et al.}, \ }%
  \bibfield{journal}{%
  \Doi{10.1088/0953-8984/20/26/264007}{\bibinfo {journal} {J. Phys.: Condens.
  Matter}}\ }%
  \textbf{\bibinfo {volume} {20}},\ \bibinfo {pages} {264007} (\bibinfo {year}
  {2009})%
  \bibAnnoteFile{NoStop}{Koster:jpcm08}%
\bibitem{Vaknin:prb89}%
  \BibitemOpen
  \bibfield{author}{%
  \bibinfo {author} {\bibfnamefont{D.}~\bibnamefont{Vaknin}}, \bibinfo {author}
  {\bibfnamefont{E.}~\bibnamefont{Caignol}}, \bibinfo {author}
  {\bibfnamefont{P.~K.}\ \bibnamefont{Davies}} {\it et al.}, \ }%
  \bibfield{journal}{%
  \Doi{10.1103/PhysRevB.39.9122}{\bibinfo {journal} {Phys. Rev. B}}\ }%
  \textbf{\bibinfo {volume} {39}},\ \bibinfo {pages} {9122} (\bibinfo {year}
  {1989})%
  \bibAnnoteFile{NoStop}{Vaknin:prb89}%
\bibitem{Goniakowski:prl07}%
  \BibitemOpen
  \bibfield{author}{%
  \bibinfo {author} {\bibfnamefont{J.}\ \bibnamefont{Goniakowski}}\ and\
  \bibinfo {author} {\bibfnamefont{C.}\ \bibnamefont{Noguera}},\ and\
  \bibinfo {author} {\bibfnamefont{L.}\ \bibnamefont{Giordano}},\ }%
  \bibfield{journal}{%
  \Doi{10.1103/PhysRevLett.98.205701}{\bibinfo {journal} {Phys. Rev. Lett.}}\ }%
  \textbf{\bibinfo {volume} {98}},\ \bibinfo {pages} {205701} (\bibinfo {year} {2007})%
  \bibAnnoteFile{NoStop}{Goniakowski:prl07}%
\bibitem{Goniakowski:prl04}%
  \BibitemOpen
  \bibfield{author}{%
  \bibinfo {author} {\bibfnamefont{J.}\ \bibnamefont{Goniakowski}}\ and\
  \bibinfo {author} {\bibfnamefont{C.}\ \bibnamefont{Noguera}},\ and\
  \bibinfo {author} {\bibfnamefont{L.}\ \bibnamefont{Giordano}},\ }%
  \bibfield{journal}{%
  \Doi{10.1103/PhysRevLett.93.215702}{\bibinfo {journal} {Phys. Rev. Lett.}}\ }%
  \textbf{\bibinfo {volume} {93}},\ \bibinfo {pages} {215702} (\bibinfo {year} {2004})%
  \bibAnnoteFile{NoStop}{Goniakowski:prl04}%
\bibitem{Freeman:prl06}%
  \BibitemOpen
  \bibfield{author}{%
  \bibinfo {author} {\bibfnamefont{C.~L.}\ \bibnamefont{Freeman}}\ and\
  \bibinfo {author} {\bibfnamefont{F}\ \bibnamefont{Claeyssens}},\ and\
  \bibinfo {author} {\bibfnamefont{N.~L}\ \bibnamefont{Allan}},\ and\
  \bibinfo {author} {\bibfnamefont{J.~H}\ \bibnamefont{Harding}},\ }%
  \bibfield{journal}{%
  \Doi{10.1103/PhysRevLett.96.066102}{\bibinfo {journal} {Phys. Rev. Lett.}}\ }%
  \textbf{\bibinfo {volume} {96}},\ \bibinfo {pages} {066102} (\bibinfo {year} {2006})%
  \bibAnnoteFile{NoStop}{Freeman:prl06}%
\bibitem{Tusche:prl07}%
  \BibitemOpen
  \bibfield{author}{%
  \bibinfo {author} {\bibfnamefont{C.}\ \bibnamefont{Tusche}}\ and\
  \bibinfo {author} {\bibfnamefont{H.~L.}\ \bibnamefont{Meyerheim}},\ and\
  \bibinfo {author} {\bibfnamefont{J.}\ \bibnamefont{Kirschner}},\ }%
  \bibfield{journal}{%
  \Doi{10.1103/PhysRevLett.99.026102}{\bibinfo {journal} {Phys. Rev. Lett.}}\ }%
  \textbf{\bibinfo {volume} {99}},\ \bibinfo {pages} {026102} (\bibinfo {year} {2007})%
  \bibAnnoteFile{NoStop}{Tusche:prl07}%
\bibitem{Pavlenko:prb07}%
  \BibitemOpen
  \bibfield{author}{%
  \bibinfo {author} {\bibfnamefont{N.}~\bibnamefont{Pavlenko}}, \bibinfo
  {author} {\bibfnamefont{I.}~\bibnamefont{Elfimov}}, \bibinfo {author}
  {\bibfnamefont{T.}~\bibnamefont{Kopp}},\ and\ \bibinfo {author}
  {\bibfnamefont{G.~A.}\ \bibnamefont{Sawatzky}},\ }%
  \bibfield{journal}{%
  \Doi{10.1103/PhysRevB.75.140512}{\bibinfo {journal} {Phys. Rev. B}}\ }%
  \textbf{\bibinfo {volume} {75}},\ \bibinfo {pages} {140512} (\bibinfo {year}
  {2007})%
  \bibAnnoteFile{NoStop}{Pavlenko:prb07}%
\bibitem{Note1}%
  \BibitemOpen
  \bibinfo {note} {consistent with ``uncompensated polarity'' in MgO films, see \cite{Goniakowski:prl07}}%
  \bibAnnoteFile{NoStop}{Note1}
\bibitem{Note2}%
  \BibitemOpen
  \bibinfo {note} {Theoretical studies for MgO(111) \cite{Goniakowski:prl04} and ZnO \cite{Freeman:prl06} polar thin films predict analogous changes in atomic structures driven by electrostatic instabilities. Such  changes have recently been observed \cite{Tusche:prl07}.}%
  \bibAnnoteFile{NoStop}{Note2}
\end{thebibliography}
\end{document}